\documentclass[preprint,12pt]{elsarticle}

\usepackage{bm}
\usepackage{amsmath}
\usepackage{amssymb}

\journal{Physics Letters A}
\begin{document}

\begin{frontmatter}

\title{An analytic approximation to the Diffusion Coefficient for the periodic Lorentz Gas}
\author[math]{C. Angstmann}
\ead{c.angstmann@unsw.edu.au}
\author[phys]{G. P. Morriss}
\ead{g.morriss@unsw.edu.au}
\address[phys]{School of Physics, University of New South Wales, Sydney Australia 2052}
\address[math]{School of Mathematics and Statistics, University of New South Wales, Sydney Australia 2052}

\begin{abstract}
An approximate stochastic model for the topological dynamics of the periodic triangular Lorentz gas is constructed. The model, together with an extremum principle, is used to find a closed form approximation to the diffusion coefficient as a function of the lattice spacing. This approximation is superior to the popular Machta and Zwanzig result and agrees well with a range of numerical estimates. 
\end{abstract}

\end{frontmatter}

\section{Introduction}

The diffusion coefficient is perhaps the simplest example of a transport coefficient as it describes the transport of mass in a system. 
The Lorentz gas, originally proposed as a model for the movement of electrons in a crystal lattice, comprises a single particle moving in a lattice of fixed scatterers, and in the absence of a field the electron diffuses through the lattice. 

A variety of methods have been employed to calculate the diffusion coefficient of the Lorentz gas.  
Machta and Zwanzig \cite{Machta:1983qy} used a Markov hopping process to generate an analytical diffusion coefficient approximation. 
A similar approach has been used more recently for the three-dimensional Lorentz gas \cite{GNS11}.
As the model used by Machta and Zwanzig is stochastic it will be examined in more depth in section \ref{mz}. 
Morriss and Rondoni \cite{Morriss:1994fj} used the periodic orbit expansion method to calculate the diffusion coefficient but
also performed detailed calculations using the Green-Kubo relations and the mean squared displacement. 
Baranyai, Evans, and Cohen  \cite{Baranyai:1993uq} calculated the diffusion coefficient based on the Green-Kubo relation although only for a limited number of densities. 
Gaspard and Baras \cite{Gaspard:1995vn} calculated the diffusion coefficient using an escape rate formalism which has been extended to other systems \cite{HG01}.
Although these results have been known for some time there remains much interest in diffusion in billiard systems \cite{flower} and the periodic Lorentz gas \cite{Ch97,D00,MS08,GS09}.

\section{Lorentz Gas Parameters}
\begin{figure}
\begin{center}
\includegraphics[width=0.5\textwidth]{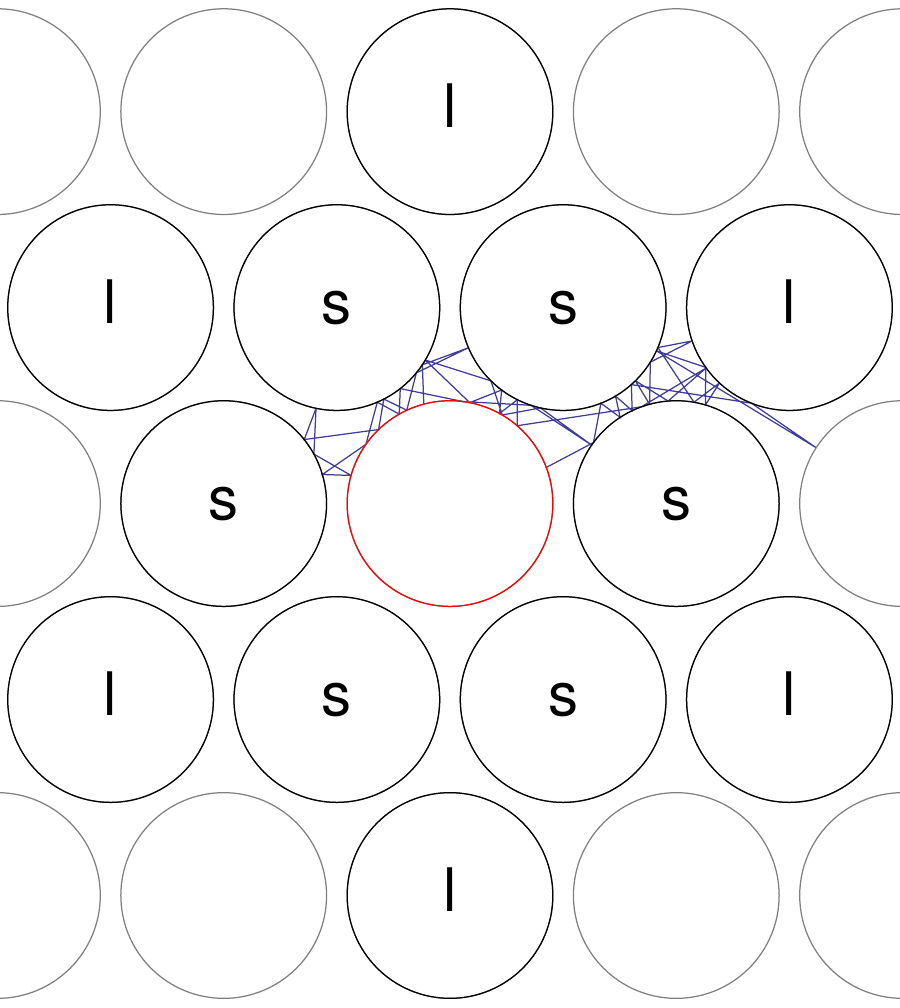}
\end{center}
\caption{\label{fig_states}A section of the periodic triangular Lorentz gas including a typical path for the wandering particle beginning from the central scatterer. This path becomes the basis for the deterministic model that we introduce later. The two symbol state space used to label each segment of trajectory is a relative symbol defined by shifting the trajectory so that the segment originates from the central scatterer, and then choosing the symbol by the next scatterer it hits. Thus a flight from the central scatterer to a nearest neighbor is labelled {\it s} for short and flight to a second nearest neighbor is labelled {\it l} for long. }
\end{figure}

We consider a triangular lattice of hard disk scatterers with a minimal spacing of $w$ between the surfaces of adjacent disks, as shown in Figure \ref{fig_states}. 
The wandering particle moves in a straight line in the area outside the scatterers until it has a specular collision with a scatterer.
As long as the spacing is small enough, $w<\frac{4}{\sqrt{3}}-2$, the horizon is finite and there are no paths of infinite length, so a collision must occur. 
The path of the wandering particle can be described by a symbolic dynamics \cite {CGS92} constructed by assigning a symbol to the next (relative) scatterer. 
If the scatterer is a nearest neighbour then we can assign that event as a {\it short} flight. If the scatterer was a next nearest neighbour then that path is a {\it long} flight (see Figure \ref{fig_states}).
Any displacement of the wandering particle in the Lorentz gas with finite horizon is composed of a combination of short and long flights. 
Although this approach has the flavour of a periodic orbit expansion, at no stage do we use periodic orbits, and the similarity is just in the symbolic dynamics used.
In the stochastic model we develop the symbolic dynamics to become the state space.

\section{\label{mz}The Machta and Zwanzig Result}

\begin{figure}
\begin{center}
\includegraphics[width=0.5\textwidth]{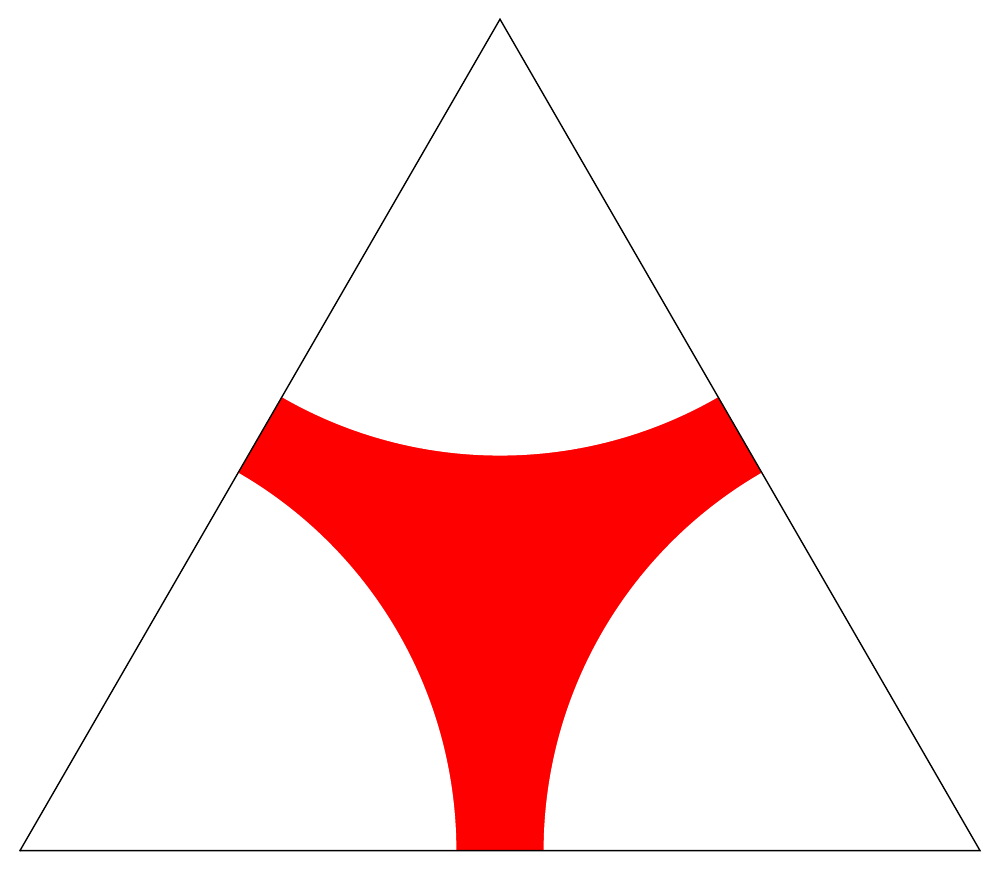}
\end{center}
\caption{\label{fig_mz}The Machta-Zwanzig trap indicated by the shaded region between three scatterers in the triangular lattice. In Figure \ref{fig_states} there is a trap between every group of three scatterers.}
\end{figure}

Machta and Zwanzig \cite{Machta:1983qy} construct their simple analytical estimate for the diffusion coefficient by replacing the trajectory of a particle with a random walk between triangular regions of the lattice called {\it traps}. A diagram of the {\it trapping} region is given in Figure \ref{fig_mz}. The probability of a transition from this region is calculated by considering the volume of phase space that will leave the trap in a time $\Delta t$. This leads to an expression for the mean occupation time of a trap as:
\begin{equation}
\tau=\frac{\pi}{6w}\left(\frac{\sqrt{3}(w+2)^{2}-\pi}{2}\right).
\end{equation}
The diffusion coefficient on a two-dimensional isotropic lattice can be expressed as:
\begin{equation}
D=\frac{l^{2}}{4\tau}
\end{equation}
where $l=(2+w)/\sqrt{3}$ is the distance between the traps. 
This gives an approximation for the diffusion coefficient as:
\begin{equation}\label{eq_mz}
D_{\mathrm{MZ}}=\frac{w(w+2)^{2}}{\pi(\sqrt{3}(w+2)^{2}-2\pi)}.
\end{equation}

This derivation relies on the assumption that the process of transitions between traps is Markov.
The more collisions that occur in the trap, the more  information about which hole the particle entered by is lost. 
The state space that Machta and Zwanzig use differs from the state space that will be employed here. 
There are similarities in the approaches but the method that is outlined here does not require the same Markov assumption. 
The Markov assumption is only justified at very small $w$ whereas the approximation outlined below should be applicable over a wider range of values.

Klages and Dellago \cite{Klages:2000lr} and Klages and Korabel \cite{KK02} have developed successive systematic refinements based on explicit correlations of the non-Markov events to improve the accuracy of the Machta and Zwanzig result.

\section{A Deterministic Model}

The diffusion coefficient is defined in terms of the linear growth of the mean-square displacement with time, and can be found using the Einstein relation:
\begin{equation}
\label{eq_diff_1}
D=\lim_{t\rightarrow\infty}\frac{\langle \Delta r^{2}(t)\rangle}{4t}.
\end{equation}
Here $\Delta {\bf r}(t)$ is the displacement of the particle as a function of time.
For the Lorentz gas the dynamics consists of repetitions of a free-flight at velocity $v=1$, followed by a collision with a scatterer giving the diffusion coefficient as a function of the spacing between the scatterers $w$.  
If no infinite length flights are possible, the Lorentz gas is said to have a {\it finite horizon} and the free-flights are of two types; flights between nearest neighour scatterers (short flights) and flights between second nearest neighbours (long flights).
Any physical trajectory can be written as a sequence of short and long flights, where each flight has a length $r_{i}$ and a unit vector direction ${\boldsymbol {\omega}}_{i}$.

To evaluate the diffusion constant from the Einstein relation we need both the length of the trajectory and the time taken to travel along it. 
The total time after $N$ flights can be written as the sum of the times for each flight $\delta t_{i}$ as 
\begin{equation}
t_{N}=\sum_{i=1}^{N}\delta t_{i} = \sum_{i=1}^{N} \frac {r_{i}} {v} = \frac {N_{s} \langle r_{s}\rangle + N_{l} \langle r_{l}\rangle} {v}
\end{equation}
where $N_{s}$ is the number of short flights, $N_{l}$ is the number of long flights, $\langle r_{s}\rangle$ is the average length of a short flight and $\langle r_{l}\rangle$ is the average length of a long flight. 
The square displacement $\Delta r^{2} (t_{N})$ at time $t_{N}$ is given by 
\begin{equation}\label{msd}
\Delta r^{2} (t_{N})=\big( \sum_{i=1}^{N} r_{i} {\boldsymbol {\omega}}_{i}\big)^{2}.
\end{equation}
Notice that despite the fact that in the Lorentz gas the flights occur in a particular order, the value of both the square displacement and the time do not depend on that ordering. 
Thus we can rearrange the order of the terms in the sum in equation \ref{msd} to collect together the short and long flights separately as
\begin{equation}
\label{delr}
\begin{split}
\Delta r^{2} (t) &= \left(  \sum_{i=1}^{N_{s}} r_{i} \boldsymbol {\omega}_{i}  + \sum_{i=1}^{N_{l}} r_{i} \boldsymbol {\omega}_{i} \right)^{2}\\
&=\left(  \sum_{i=1}^{N_{s}} r_{i} {\boldsymbol {\omega}}_{i} \right)^{2}+ \left( \sum_{j=1}^{N_{l}} r_{j} {\boldsymbol {\omega}}_{j} \right)^{2}
+ 2 \left( \sum_{i=1}^{N_{s}}r_{i}{\boldsymbol {\omega}}_{i} \right) \cdot \left( \sum_{j=1}^{N_{l}}r_{j}{\boldsymbol {\omega}}_{j} \right) .\\
&= \sum_{i=1}^{N_{s}} r_{i}^{2} + \sum_{i \neq j}^{N_{s}} r_{i} r_{j} {\boldsymbol {\omega}}_{i} \cdot {\boldsymbol {\omega}}_{j} 
+  \sum_{j=1}^{N_{l}} r_{j}^{2} +  \sum_{i \neq j}^{N_{l}} r_{i} r_{j} {\boldsymbol {\omega}}_{i} \cdot {\boldsymbol {\omega}}_{j}
+ 2  \sum_{i=1}^{N_{s}} \sum_{j=1}^{N_{l}} r_{i} r_{j} {\boldsymbol {\omega}}_{i} \cdot  {\boldsymbol {\omega}}_{j}  .\\
\end{split}
\end{equation}

 Averaging the square displacement over all initial conditions, assuming that $r_{i}$ and ${\boldsymbol {\omega}}_{i}$ are uncorrelated and that the average $\langle {\boldsymbol {\omega}}_{i} \cdot {\boldsymbol {\omega}}_{j}\rangle = 0$  for all $i \neq j$, we obtain
 
\begin{equation}
\label{delr}
\begin{split}
\langle \Delta r^{2} (t) \rangle = N_{s} \langle r_{s}^{2} \rangle +  N_{l} \langle r_{l}^{2} \rangle
\end{split}
\end{equation} 
The assumption that $\langle {\boldsymbol {\omega}}_{i} \cdot {\boldsymbol {\omega}}_{j}\rangle = 0$ is only likely to miss correlations between subsequent events, especially when both events are short flights.
The dominant term in repeated short flights will be when the two flights are in nearly opposite directions, and then  $\langle {\boldsymbol {\omega}}_{i} \cdot {\boldsymbol {\omega}}_{j}\rangle < 0$.
Repeated long flights will also produce negative correlations but these occurrences are much rarer.

   The probability of a short flight is $P_{s} = N_{s}/N$ and the probability of a long flight is $P_{l} = N_{l}/N$ so that in the limit $N\rightarrow\infty$ the diffusion coefficient becomes
\begin{equation}
\label{eq_diff_1a}
D= \frac{P_{s} \langle r^{2}_{s}\rangle + P_{l} \langle r^{2}_{l}\rangle}{4(P_{s} \langle r_{s}\rangle + P_{l} \langle r_{l}\rangle)}
\end{equation}

To evaluate this expression for the diffusion coefficient we need values for the averages $\langle r_{s}\rangle$, $\langle r_{l}\rangle$, $\langle r^{2}_{s}\rangle$ and $\langle r^{2}_{l}\rangle$, and the values of the probabilities $P_{s}$ and $P_{l}=1-P_{s}$ which will all be functions of the spacing $w$. 
The physical constraints on the system set lower bounds on $r_{s}$ and $r_{l}$ (and hence on their averages), as they cannot be lower than the minimum separation between the relevant scatterers.
Thus $r_{s}>w$ and $r_{l}>\sqrt{3}(w+2)-2$.
The values of the averages could be found numerically or by explicit integration over the billiard measure, as could the probabilities $P_{s}$ and $P_{l}$, but it is less difficult to calculate the diffusion coefficient numerically.

   Equation \ref{eq_diff_1a} gives an upper bound on the diffusion coefficient as the correlations that are ignored are mostly negative and in the numerator. 
   To gain some quantitative understanding  of the accuracy it is perhaps interesting to consider the numerical values obtain for various quantities at a single value of the spacing, $w=0.2$.
   Here the deterministic model (Equation \ref{eq_diff_1a}) gives $D=0.208$ rather than the correct value $D=0.17$. This is obtained using the average values $\langle r_{s}\rangle = 0.446$,  $\langle r_{l}\rangle=1.88$, $\langle r^{2}_{s}\rangle=0.2576$ and $\langle r^{2}_{l}\rangle=3.56$.

\section{Stochastic Model}

Rather than proceeding with the deterministic model of the previous section we will use its structure, in particular Equation (\ref{eq_diff_1a}), to construct a stochastic model for the deterministic Lorentz gas. 
 The probability space for this model will be formed from the phase space of the Lorentz gas together with its natural measure. Rather than considering the whole state space we will take the Poincar\'e section and develop a discrete model comprised of the two states, long flights and short flights. The state space labels transitions between iterations on the Poincar\'e surface rather than parameterising the position on the surface. 
The stochastic model constructed in this fashion will not be Markovian, as the correlations in the deterministic trajectory persist but this is not an issue as only the long run, or stationary, probabilities are required. From this it is obvious that  all transient behaviour is lost.  

   Our model for the diffusion coefficient   is given by
\begin{equation}
\label{eq_diff_1s}
D= \frac{P_{s} \hat r^{2}_{s} + P_{l} \hat r^{2}_{l}}{4(P_{s} \hat r_{s} + P_{l} \hat r_{l})}
\end{equation}
where the  $P_{s}$ and $P_{l} = 1 - P_{s}$ are the probabilities for short flights and long flights. 
   Here $\hat r_{s}$ and $\hat r_{l}$ are considered to be parameters which may depend on the spacing $w$.
   There is now no mapping of this stochastic model on to random walk on a lattice as the determinism of the previous model has been lost.
   
As we still have three parameters in the stochastic model, we look for a method of reducing this number.
To do this we consider  looking for an extremum of $D$ as a function of the parameters $\hat r_{s}$ and $\hat r_{l}$. 
If we minimise $D$ with respect to both $\hat r_{l}$ and $\hat r_{s}$ we find that the required value of $\hat r_{l}$ is less than the physical minimum for $r_{l}$, that is $\sqrt{3}(w+2)-2$. 
So we set  $\hat r_{l}$ at this lower bound and minimise $D$ with respect to $\hat r_{s}$ alone.



The calculation of the extremum is straightforward by differentiating equation \ref{eq_diff_1a} with respect to $\hat r_{s}$, we find that
\begin{equation}
\frac{\partial D}{\partial \hat r_{s}} = \frac{\hat r_{s}^{2} P_{s}^{2}+P_{s} P_{l} \hat r_{l} (2 \hat r_{s} - \hat r_{l})}{4(\hat r_{s}P_{s}+\hat r_{l}P_{l})}
\end{equation}
which we set to zero. The positive solution of the quadratic in $\hat r_{s}$  gives a minimum at
\begin{equation}
\hat r_{s} = r^{*}_{s}=\frac{\hat r_{l}(\sqrt{P_{l}}-P_{l})}{P_{s}}
\end{equation}
Notice that $r^{*}_{s}$ now depends on $w$ through the dependence of the probabilities $P_{s}$ and $P_{l}$ and the fixed value of $\hat r_{l}$. 
The stochastic model now only depends upon a single parameter $P_{l}=1-P_{s}$. 
After some algebraic manipulation the diffusion coefficient is given by
\begin{equation}
\label{eq_dp}
D=\frac{\hat r_{l}(\sqrt{P_{l}}-P_{l})}{2 P_{s}}=\frac{r^{*}_{s}}{2} .
\end{equation}

\section{Probabilities}

\begin{figure}
\includegraphics[width=1\textwidth]{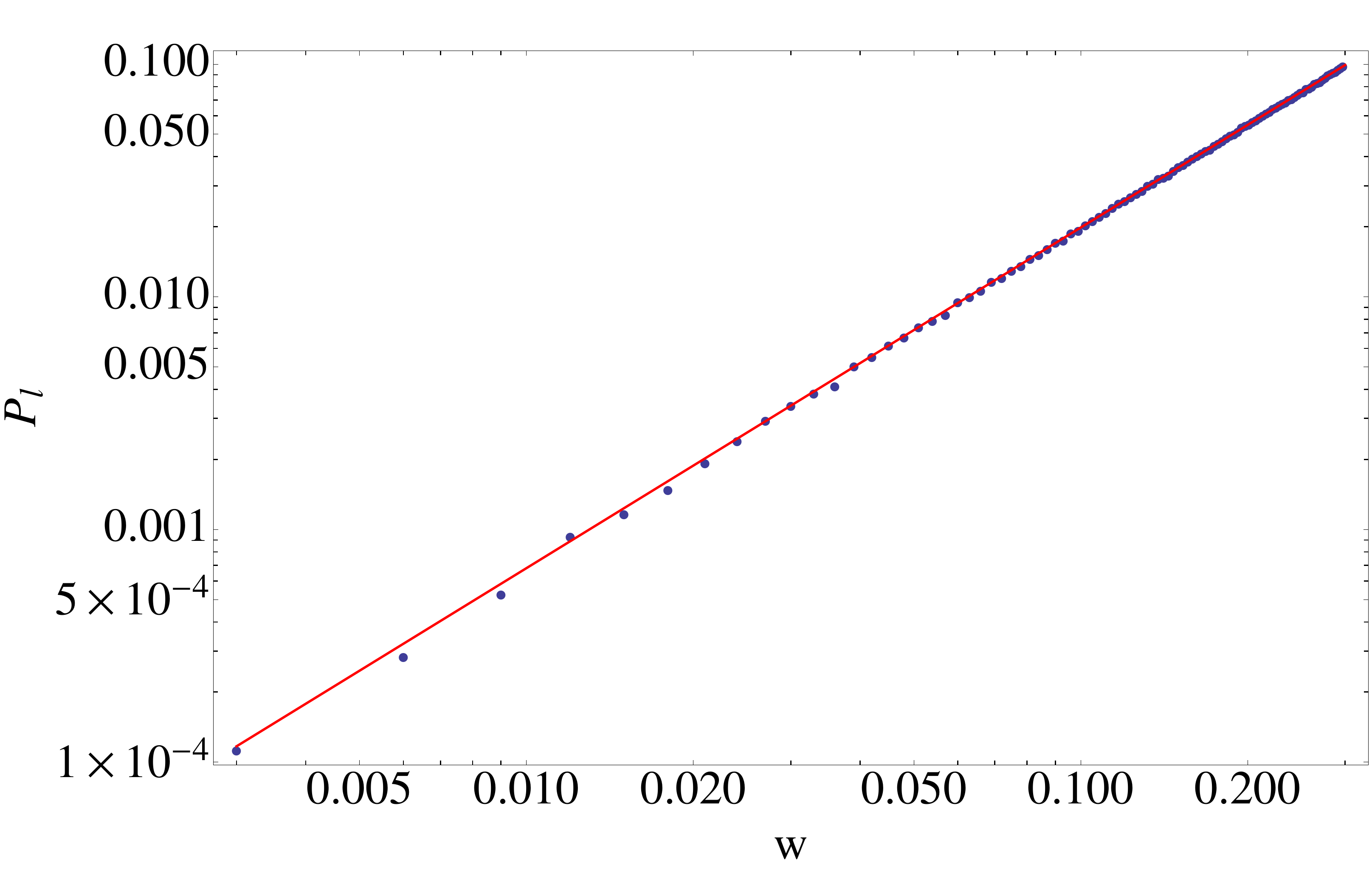}
\caption{The probability of a long flight as a function of scatterer spacing $w$. Each point is a molecular dynamics calculation of at least $5\times 10^5$ collisions and the solid line is the power law fit $P_{l}=\beta w^{\alpha}$. Notice that this fit is a very good approximation throughout the whole range of spacings. }
\label{prob_lf}
\end{figure}

The probabilities of short and long flights should be an integral over the regions of initial conditions, weighted with the uniform measure, leading to short and long flights and hence are in principal exactly computable. 
Rather than undertake this computation directly, the probability is calculated from a numerical simulation which is equivalent to a Monte-Carlo evaluation of the integral. 
 We calculate  the probability based on the frequency of occurrence of short and long flights over a range of spacings $w$ for trajectories of at least $5 \times 10^5$ collisions. The probability of a long flight $P_{l}$ as a function of $w$ appears continuous and smooth and is fitted well by a simple power law in $w$. 
\begin{equation}
\label{eq_plf}
P_{l}=\beta w^{\alpha}
\end{equation}
where the coefficient choices $\alpha=2 ( \sqrt{3}-1)$ and $\beta =\frac{1}{\sqrt{3}}$, give a good fit to the data.
We plot the simulation probabilities and the power law fit in Figure \ref{prob_lf}.   

\begin{figure}
\includegraphics[width=1\textwidth]{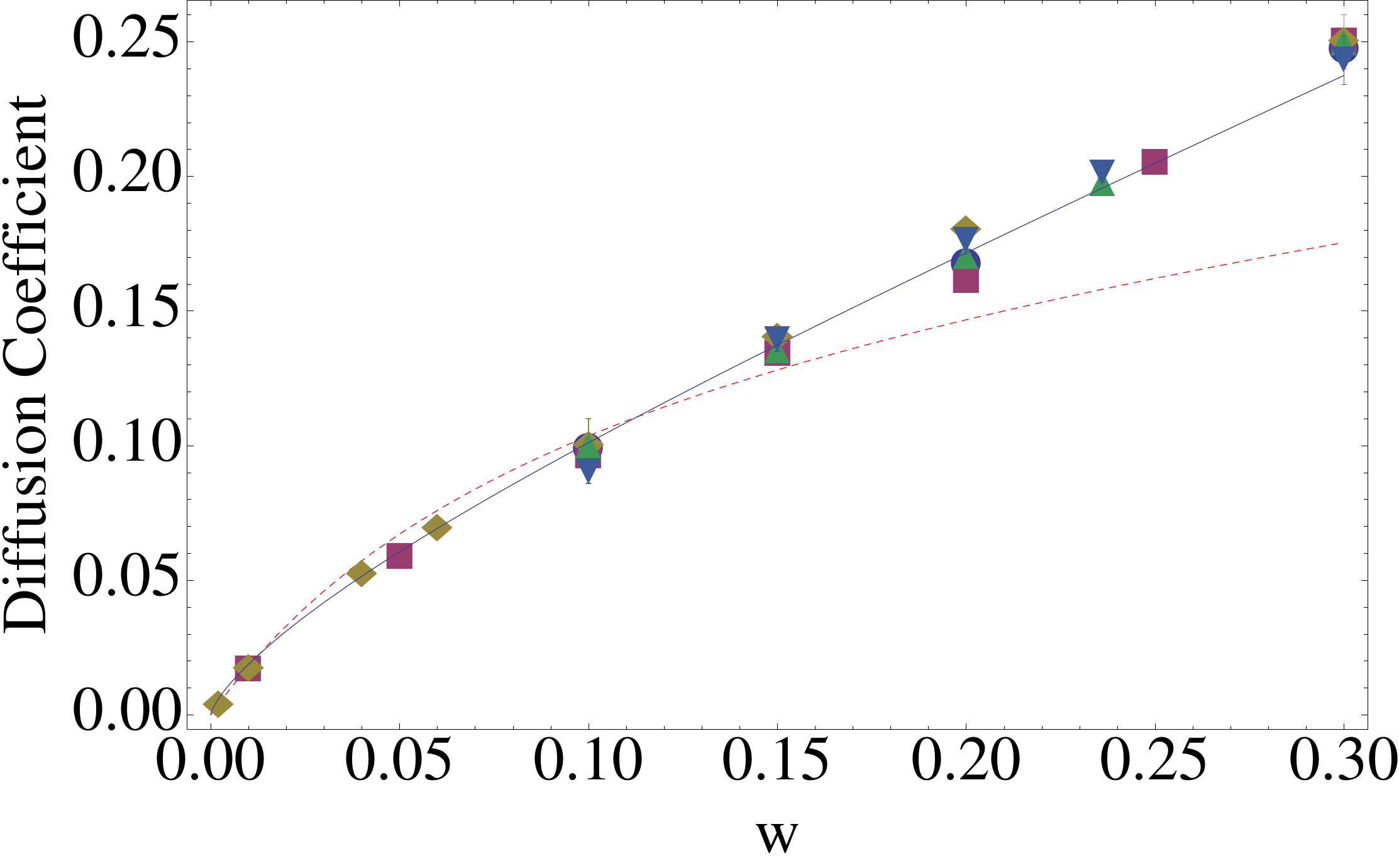}

\caption{A comparison of values of the diffusion coefficient obtained from numerical simulation with two analytic results, that of Machta and Zwanzig (Equation \ref{eq_mz} the dashed line), and the result obtained here in Equation \ref{D_coeff} (the solid line). The diamonds are Green Kubo results from  \cite{Machta:1983qy}, the squares are escape rate calculations from Gaspard and Baras \cite{Gaspard:1995vn}, and the filled circles are Green Kubo calculations from Baranyai, Evans, and Cohen  \cite{Baranyai:1993uq}. The triangles are the Green Kubo results, and the upside down triangles are periodic orbit calculations from Morriss and Rondoni \cite{Morriss:1994fj}.}
\label{dif_min}
\end{figure}

  Numerical results for the diffusion coefficient are obtained using the fitted expression for the probability $P_{l}$ and the lower bound for $\hat r_{l}$, thus the final result for the diffusion coefficient is
\begin{equation}
\label{D_coeff}
D=\frac { \left(\sqrt{3} (w+2) - 2 \right) \left( \sqrt{\beta} w^{\alpha/2}-\beta w^{\alpha} \right) } {2 \left( 1 - \beta w^{\alpha} \right) }
\end{equation}
A comparison of these results with simulation results is given in figure \ref{dif_min}. This equation fits the other data remarkably well, only falling outside the error bars at $w=0.3$ and for very small values of $w$. A deviation may be expected at large $w$ as we are using the lower bound  for $r_{l}$. 

    At $w=0.2$ the stochastic model (Equation \ref{D_coeff}) gives a value of $D=0.171$, using $r^{*}_{s}=0.343$ which less than the value of $\langle r_{s}\rangle$ and should therefore lead to a smaller value for $D$.
   At first glance, the stochastic model appears to replace $\langle r^{2}_{s}\rangle$ by $\langle r_{s}\rangle^{2}$ and the same replacement for the long flights. 
   The difference in these two terms is $\langle r^{2}_{s}\rangle - \langle r_{s}\rangle^{2} =0.058$ and $\langle r^{2}_{l}\rangle - \langle r_{l}\rangle^{2} =0.03$ which accounts for only part of the difference between the diffusion coefficients of the deterministic and stochastic models.

The fact that the minimum value of the diffusion coefficient gives the best value for $D$ over the whole range of $w$ suggests that there is some overall minimum principle at work, although there is no strong justification for this.

\section{Small $w$ limit}


The limiting behaviour of the diffusion coefficient is incorrect in the small $w$ limit, where linear behaviour is expected  \cite{Bunimovich:1985yq}. 
 For small $w$  the power law fit for the probability, Equation \ref{eq_plf}, may be less accurate. 
The derivation of the diffusion coefficient is independent of the form of the probability. For the derivation to hold and the diffusion coefficient to be linear as $w\rightarrow0$ the probability must be of the form:
\begin{equation}
P_{l}\approx a w^{2}
\end{equation}
where $a$ is the slope approaching $w=0$. 
   Taking the  following form for the probability we can ind a better fit over the entire range of $w$:
\begin{equation}
P_{l}= \beta (e^{-\gamma w} w^{2}+(1-e^{-\gamma w})w^{\alpha})
\end{equation}
where $\alpha$ and $\beta$ are as given before. The parameter $\gamma$ is fitted to the simulation data and found as $167.37$. With this probability the diffusion coefficient has the correct limiting behaviour for small $w$. 

\section{Conclusion}

   The simple stochastic model presented here works remarkably well over the full range of physical $w$ values. Based on very little information, the diffusion coefficient for the Lorentz gas in Equation \ref{D_coeff} appears in agreement with previous numerical results. 
   The main advantage of the present method is computational simplicity and surprising accuracy. 
   A dynamically motivated stochastic model, combined with a minimisation procedure and a simple power law approximation to the  probabilities gives an accurate closed form approximation to the diffusion coefficient. 
   It can be corrected at very small $w$ to have the correct limiting behaviour.
   
\bibliographystyle{model1-num-names}
\bibliography{LGD}

\end{document}